\documentclass[twocolumn]{article}

\usepackage{graphicx,times}             
\usepackage{natbib}
\usepackage{amssymb,amsmath}
\bibpunct{(}{)}{;}{a}{}{,}

\usepackage[cm]{fullpage}
\usepackage{bm}
\usepackage[version=4]{mhchem}
\usepackage{xcolor,listings}
\usepackage{hyperref}
\usepackage{aa}
\usepackage{fullpage}

\newcommand\code[1]{\textsl{#1}}
\newcommand\codename{\textsc{Chempl}}
\newcommand\CODENAME{\textsc{Chempl}}
\newcommand\codenameTT{{Chempl}}
\newcommand\set[1]{\mathcal{#1}}
\newcommand\refsec[1]{Section~\ref{#1}}
\newcommand\reffig[1]{Fig.~\ref{#1}}
\newcommand\refeq[1]{equation~(\ref{#1})}
\newcommand\reftab[1]{Table~\ref{#1}}
\newcommand\refstep[1]{step~\ref{#1}}
\newcommand\ssd{\rho_\text{S}}
\newcommand\NSS{N_\text{S}}
\newcommand\ns[1]{n_\text{S}(#1)}
\newcommand\nga[1]{n_\text{G}(#1)}
\newcommand\nm[1]{n_\text{M}(#1)}
\newcommand\rtlv{RATE12}
\newcommand\vc[1]{\bm{#1}}

\begin{document}

\title{\codenameTT{}: a playable package for modeling interstellar chemistry}

\author{Fujun Du$^{1,2}$}
\date{\small%
$^1$Purple Mountain Observatory and Key Laboratory of Radio Astronomy, Chinese Academy of Sciences, Nanjing 210023, P.R.China; \textbf{fjdu@pmo.ac.cn} \\
$^2$School of Astronomy and Space Science, University of Science and Technology of China, Hefei 230026, P.R.China}

\maketitle

\begin{abstract}
Astrochemical modeling is needed for understanding the formation and evolution of interstellar molecules, and for extracting physical information from spectroscopic observations of interstellar clouds.  The modeling usually involves handling of a chemical reaction network and solution of a set of coupled nonlinear ordinary differential equations, which is traditionally done using code written in compiled languages such as Fortran or C/C++.  While being computationally efficient, there is room for improvement in the ease of use and interactivity for such an approach.  In this work we present a new public code named \codename{}, which emphasizes interactivity in a modern Python environment, while remaining computationally efficient.  Common reaction mechanisms and a three-phase formulation of gas-grain chemistry are implemented by default.  It is straightforward to run 0D models with \codename{}, and only a small amount of additional code is needed to construct 1D or higher dimensional chemical models.  We demonstrate its usage with a few astrochemically relevant examples.
\end{abstract}


%
%
\section{Introduction}           
\label{secIntro}

The interstellar medium (ISM) is ever-evolving under the effect of electromagnetic radiation, cosmic-rays, and hydrodynamic motion coupled with gravitational and magnetic field.  As the physical condition changes, the chemical composition changes accordingly.  The buildup of chemical complexity in the ISM, the chemical heritage of forming stellar and planetary systems, and the possible connection between astronomy and the origin of life are fascinating topics.  Over the years the chemistry of ISM has proved to be a powerful diagnostic of the physical parameters.  For this to work at a quantitative level, it is necessary to model the chemistry in detail.  Except for simple cases, a comprehensive chemical network has to be used, because most of the time it is not easy to figure out beforehand which reactions are important and which are not.  With over 200 molecules detected in space (not counting isotopologues) \citep{Endres2016,McGuire2018}, the scale of chemical networks for modeling interstellar molecules keeps growing, with the largest at the moment containing tens of thousands of reactions.  The evolution of a large and interconnected chemical network can only be tracked with numerical methods.

In the field of astrochemistry --- namely, the study of the chemical side of astronomy --- there are already a few codes, many of which are publicly available, including:
Nahoon \citep{Wakelam2012},
MAGICKAL \citep{Garrod2013},
KROME \citep{Grassi2014},
Astrochem \citep{Maret2015},
Nautilus \citep{Ruaud2016},
ALCHEMIC \citep{Semenov2017}, and
UCLCHEM \citep{Holdship2017}.  So why create yet another code?

Traditionally, codes for astrochemistry are written in Fortran or C/C++, and the mode of running is usually like this: (1) The user prepares a few files containing the reaction network, the initial chemical abundances, and the physical parameters, (2) then feeds these files to the compiled program to get the evolution of the abundances of different species as a function of time, with the results saved in files.  In some codes a preprocessing step to generate the source code describing the differential equations from a reaction network is needed.  (3) To analyze the results, separate code has to be written to load the files containing the results.  While this whole process can be pipelined and parallelized (at the process level) to work efficiently, it lacks interactivity.  The user has to switch between different execution environments back and forth, e.g., a command line terminal to run the chemical code, and a visualization environment (such as Python/matplotlib) for analyzing and making plots.  Such a context switching and the associated data loading could be cumbersome.  When working with models with time-dependent physical conditions, one might have to hard-code the time-dependency into the code and recompile the binary executable each time a change is made (it is noted that Nautilus uses input files to implement time-dependency), which could be distracting to say the least.

Here we present a new code \codename{}\footnote{\codename{} is available at \url{https://github.com/fjdu/chempl}.  A few reaction networks and initial conditions are also included in this repository.  All the scripts for the example models in this paper are available at \url{https://github.com/fjdu/chempl/blob/master/Examples-for-the-chempl-paper-2020-05.ipynb}.}, which is aimed at improving interactivity without sacrificing efficiency.  In writing \codename{}, one goal we keep in mind is to make the code as ``playable'' as possible, hence the name.  It should be relatively straightforward to make changes to the model configuration, and be fast to see the effects of the changes that just have been made.
The user is able to pause the calculation temporarily, look into the internal states of the chemical system and possibly make some changes to the physical/chemical parameters, and continue the solution afterwards.
To ensure performance, under the hood the code is written in C++, and the ODE (ordinary differential equation) solver being used is the time-tested DLSODES solver of the ODEPACK package written in Fortran \citep{Hindmarsh1983}.  The interactivity is achieved through integration with Python, and is designed to be used in the iPython or Jupyter environment.  We use Cython \citep{Behnel2011} to wrap the C++ code into a Python module.  Most of the internal model data can be accessed as a normal Python variable, and can be saved as human-readable or binary data files, or it may be more beneficial to save the Python data structures using the ``pickle'' facility of Python (or other structured data format such as the Advanced Scientific Data Format (ASDF)).  Note that \codename{} can also be used in the ``traditional'' mode (i.e., working with configuration files and launched in the command line), or incorporated into other simulations as a subroutine with a few lines of additional code for setting it up.

\codename{} by default supports common first order and second order reaction types, includes approximate treatment for the self- and mutual- shielding of H$_2$ and CO molecules, and works with comprehensive gas-grain chemical networks modeled as a three-phase system.
As will be shown in example models, it is straightforward to construct 0D or 1D chemical models with \codename{}.
Self-consistent temperature calculation based on heating and cooling balance is not yet included at the moment, but it is in the plan for the near future.
For grain surface chemistry, \codename{} currently adopts the rate equation approach as the mathematical formulation for describing the chemical dynamics.  It is known that in some physical conditions the rate equation approach does not accurately capture the stochastic features of the grain surface system, and stochastic simulation \citep{Gillespie1976,Vasyunin2009,Chang2014}, modifications to the rate equations \citep{Garrod2008}, or moment equation methods \citep{Du2011} need to be used.  Inclusion of these alternative formulations are also in the plan.

This paper focuses on code capabilities and demonstration of use cases for \codename{}, instead of the scientific implications of the models.
In \refsec{secDescription} we depict the mathematical framework for interstellar chemistry.
In \refsec{secExample} we show a few example astrochemical models simulated with \codename{}.
We conclude in \refsec{secConclusion}.

\section{Mathematical formulation of Interstellar Chemistry}
\label{secDescription}

In astrochemical modeling the abundances of different chemical species are solved as a function of time.
The differential equation (actually a set of coupled equations) to be solved is of the form
\begin{equation}
  \frac{d \vc{y}}{d t} = \vc{f}(t,\vc{y}),
  \label{eqGeneric}
\end{equation}
where $\vc{y}$ is a vector with component $y_i$  being the abundance of the $i$th species, usually measured relative to hydrogen nuclei.  The vector-valued function $\vc{f}$ on the right hand side is a nonlinear function of the components of $\vc{y}$, and may be explicitly time-dependent.

Sometimes a steady-state solution is needed, which can be obtained by solving the algebraic equation $\vc{f}(t,\vc{y}) = \vc{0}$, or by evolving $\vc{y}(t)$ for a very long time by solving \refeq{eqGeneric}.

The system of ODE equations in \refeq{eqGeneric} is often stiff, meaning that many vastly different timescales are involved, requiring that implicit, rather than explicit methods, be used for efficient solution.  Solvers of stiff ODEs often require knowledge of the Jacobian matrix of the right hand side, defined as
\begin{equation}
  J_{ij} \equiv \frac{\partial f_i}{\partial y_j},
\end{equation}
which is often sparse for astrochemistry-relevant networks, meaning that a large fraction (${\gtrsim}$90\%) of the elements of $J$ are identically zero.  \codename{} constructs the sparse structure from the reaction network.

\subsection{The canonical rate equations}

In interstellar or protoplanetary conditions, where the majority of the gas under consideration has a density typically in the range of ten to the power of a few per cm$^3$, and densities higher than ${\sim}10^{13}$~cm$^{-3}$ rarely need to be considered, three-body chemical reactions are not important and are excluded in astrochemical networks.

When only first and second order reactions are included, \refeq{eqGeneric} can be written in component form as
\begin{equation}
\begin{split}
  \frac{d y_i}{d t} = & \sum_{j\in \set{F}_{i,1}} k_j y_{j,1} + \sum_{j\in \set{F}_{i,2}} k_j y_{j,1} y_{j,2} - \sum_{j\in \set{D}_{i,1}} k_j y_{j,1} \\
  & - \sum_{j\in \set{D}_{i,2}} k_j y_{j,1} y_{j,2},
\end{split}
  \label{eqCanonical}
\end{equation}
where $\set{F}_{i,1}$ and $\set{F}_{i,2}$ are the set of first and second order reactions that form species $i$, $\set{D}_{i,1}$ and $\set{D}_{i,2}$ are the set of first and second order reactions that destroy species $i$, $k_j$ is the rate coefficient of reaction $j$, and $y_{j,1}$ and $y_{j,2}$ are the abundances of the first and second (when applicable) reactants of reaction $j$.  Note that for $j\in\set{D}_{i,1}$, $y_{j,1}$ is always simply $y_i$, and for $j\in\set{D}_{i,2}$, one of $y_{j,1}$ and $y_{j,2}$ (or both) must be the same as $y_i$.

Examples of first order reactions are photodissociation, cosmic-ray ionization, adsorption, and desorption reactions.  These reactions are not strictly ``uni-molecular'', since when one of these reactions happen there must be two particles interacting, with one being the chemical species we care about and the other being a photon, a cosmic-ray particle, or a dust grain.  They are treated as first order because the fluxes of photons and cosmic-rays are external inputs and are independent of local conditions, and in the case of adsorption the dust grains can be ``reused'' (hence effectively being constant).
Examples of second order reactions are neutral-neutral, neutral-ion, dissociative recombination, and radiative association reactions.  One may consult \citet{Wakelam2012} or \citet{McElroy2013} for a list of possible types of gas-phase reactions relevant for astrochemistry.

\subsection{Expressions for rates of different types of reactions}

In this section we describe formulae for calculating the rates for different types of reactions.  While these equations can be found in many modeling papers in astrochemistry, they are presented here for reference.  More detailed pedagogical discussions can be found in \citet{Du2012}.  For some reaction types, more realistic expressions for their rates may need to be used when modeling some specific problems.  In \codename{} it is possible to add new reaction types with associated handlers to accomplish this.

\subsubsection{Cosmic-ray reactions}

For ionization directly caused by cosmic-rays, the rate coefficient is simply a constant:
\begin{equation}
  k = \alpha\; \text{s}^{-1},
\end{equation}
where $\alpha$ takes different value for different species.
For cosmic-ray-induced photoreactions, the rate coefficient is
\begin{equation}
  k = \alpha \left(\frac{T}{300\;\text{K}}\right)^\beta \frac{\gamma}{1-\omega}\;\text{s}^{-1},
\end{equation}
where $\alpha$, $\beta$, and $\gamma$ are constants, and $\omega$ is the albedo of dust grains to dissociative photons, to which we assign a value of 0.6 by default.  Both direct and induced reaction rates due to cosmic-rays can be rescaled by the actual cosmic-ray ionization rate (relative to the fiducial rate of $\zeta_0=1.36\times10^{-17}\;\text{s}^{-1}$, see \citealt{McElroy2013}).

\subsubsection{Photoreactions}

For photoreactions (photodissociation and photoionization), the rate coefficient is
\begin{equation}
  k = \alpha \exp\left(-\gamma A_\text{V}\right)\;\text{s}^{-1},
  \label{eqSimplePhoto}
\end{equation}
where $\alpha$ is the unattenuated rate, $\gamma$ is a constant to scale the visual extinction $A_\text{V}$ to the UV band.  These rates can also be rescaled by a $G_0$ parameter, namely the ratio between the actual UV field and the Draine interstellar field \citep{Draine1978}.

For species such as \ce{H2}, CO, and \ce{N2}, self- and mutual- shielding can be important, and \refeq{eqSimplePhoto} cannot be used.  Approximate analytical formulae or tabulated rates (e.g. \citealt{Draine1996}, \citealt{Morris1983}, \citealt{Visser2009}, \citealt{Li2013}) with dependencies on column densities and temperature are available in this case.

\subsubsection{Generic two-body reaction rates}

The modified Arrhenius equation is often used to calculate the two-body reaction rate coefficients:
\begin{equation}
  k = \alpha \left(\frac{T}{300\;\text{K}}\right)^\beta \exp\left(\frac{-\gamma}{T}\right)\;\text{cm}^3\;\text{s}^{-1},
\end{equation}
where $\alpha$, $\beta$, and $\gamma$ are constants provided by the reaction network databases.
The reaction rate, or stated more clearly, the number of occurrence of a reaction per unit time per unit volume, can be expressed as
\begin{equation}
  k n_i n_j,
\end{equation}
where $n_i$ and $n_j$ are the volume density (in cm$^{-3}$) of the two reactants of a two-body reaction.  The time derivative of the density of species $i$ caused by this reaction is then (assuming $i\ne j$)
\begin{equation}
  \frac{d n_i}{d t} = - k n_i n_j.
\end{equation}
In astrochemistry the abundance of a species relative to hydrogen nuclei is usually used, i.e., $n_i = y_i n_\text{H}$, and the time derivative of $y_i$ is then
\begin{equation}
  \frac{d y_i}{d t} = -k n_\text{H}\cdot y_i y_j.
\end{equation}
Note that the modified Arrhenius equation is an empirical formula and does not apply for all reactions for all temperature ranges (see \refsec{secContReacRates}).

\subsubsection{Adsorption of gas phase species onto the surfaces of dust grains}

The number of particles of species $i$ adsorbed onto dust grains per unit volume per unit time is \citep{Du2014}
\begin{equation}
  s\,\sigma v_\text{T}(i) n_\text{dust} n_i,
\end{equation}
where $s$ is the sticking coefficient, $\sigma=\pi a^2$ is the dust cross section with $a$ being the dust grain radius, $n_\text{dust}$ and $n_i$ are the volume density of dust grains and species $i$, and $v_\text{T}(i)$ is the thermal speed of species $i$:
\begin{equation}
  v_\text{T}(i) = \left(\frac{8k_\text{B}T}{\pi m_i}\right)^{1/2},
\end{equation}
where $k_\text{B}$ is the Boltzmann constant, and $m_i$ is the mass of a particle of species $i$.
Written in terms of abundance relative to hydrogen nuclei, the time derivative is
\begin{equation}
  \frac{d y_i}{d t} = -s\,\sigma v_\text{T}(i) n_\text{dust} y_i = -s\,\sigma v_\text{T}(i) n_\text{H} \eta y_i,
\end{equation}
where $\eta$ is the dust-to-hydrogen number ratio, which can be further expressed as:
\begin{equation}
\begin{split}
  \eta &= \eta_\text{M}\cdot \frac{\mu\,m_\text{H}}{\frac{4\pi}{3}a^3\rho_\text{d}}\\
  &= 2.8{\times}10^{-12}
  \left(\frac{\eta_\text{M}}{0.01}\right)
  \left(\frac{\mu}{1.4}\right)
  \left(\frac{a}{0.1\;\mu\text{m}}\right)^{-3}
  \left(\frac{\rho_\text{d}}{2\,\text{g\,cm}^{-3}}\right)^{-1},
  \label{eqEta}
\end{split} 
\end{equation}
where $\eta_\text{M}$ is the dust-to-grain mass ratio, with a canonical value of 0.01, $m_\text{H}$ is the mass of a hydrogen nucleus, $\mu$ is the mean molecular mass per hydrogen nucleus, and $\rho_\text{d}$ is the material density of dust grains.
So the rate coefficient of adsorption reaction is
\begin{equation}
\begin{split}
  k &= s\, \pi a^2 \left(\frac{8k_\text{B}T}{\pi m_i}\right)^{1/2} n_\text{H}\, \eta_\text{M}\cdot \frac{\mu\,m_\text{H}}{\frac{4\pi}{3}a^3\rho_\text{d}} \\
  &= s\, \frac{3}{4a} \left(\frac{8k_\text{B}T}{\pi m_i}\right)^{1/2} n_\text{H} \eta_\text{M} \frac{\mu\,m_\text{H}}{\rho_\text{d}} \\
  &= 4.03{\times}10^{-12}\;\text{s}^{-1}\;s\,
     \left(\frac{a}{0.1\,\mu\text{m}}\right)^{-1}
     \left(\frac{T}{10\,\text{K}}\right)^{1/2}
     \left(\frac{m_i}{m_\text{H}}\right)^{-1/2} \\
  &\quad\times
     \left(\frac{n_\text{H}}{10^5\,\text{cm}^{-3}}\right)
     \left(\frac{\eta_\text{M}}{0.01}\right)
     \left(\frac{\mu}{1.4}\right)
     \left(\frac{\rho_\text{d}}{2\,\text{g}\,\text{cm}^{-3}}\right)^{-1}.
\end{split}
\end{equation}
The sticking coefficient $s$ can be determined by experiments and is usually in the range 0.1--1.
Note that for CO molecule, the above expression gives a depletion timescale of ${\sim}4{\times}10^4$ yr for the fiducial physical parameters in the parentheses.

\subsubsection{Thermal desorption from dust grain surfaces}

The thermal desorption rate is
\begin{equation}
  k = \nu_i \exp\left(-E_i/T_\text{d}\right),
  \label{eqThermalDesorption}
\end{equation} 
where $T_\text{d}$ is the dust temperature (which may be different from the gas temperature), and $\nu_i$ is the vibrational frequency of species $i$ on dust grain surface:
\begin{equation}
\begin{split}
  \nu_i &= \left(\frac{2\ssd{}E_i}{\pi^2 m_i}\right)^{1/2}\\
  &= 4.09{\times}10^{12} \;\text{Hz}\;
  \left(\frac{\ssd}{10^{15}\;\text{cm}^{-2}}\right)^{1/2}\\
  &\quad \left(\frac{E_i}{10^3\;\text{K}}\right)^{1/2}
  \left(\frac{m_i}{m_\text{H}}\right)^{-1/2}.
\end{split} 
\label{eqVibFreq}
\end{equation} 
$E_i$ is the desorption energy of species $i$, and $\ssd$ is the surface site density typically taken to be $10^{15}$ cm$^{-2}$.

\subsubsection{Cosmic-ray desorption}

When energetic cosmic-ray particles hit a dust grain, the latter will be promptly heated, leading to evaporation of surface species.  The evaporation only last a very brief period, because the dust grain cools down quickly.  This cosmic-ray-induced desorption is usually implemented using the formulation of \citet{Hasegawa1993a}:
\begin{equation}
  k_\text{CR} = k_i(70\,\text{K})\,f(70\,\text{K}),
\end{equation} 
where $k_i(70\,\text{K})$ is the thermal desorption rate of species $i$ calculated with \refeq{eqThermalDesorption} for a dust grain temperature of 70 K, and $f(70\,\text{K})$ is the fraction of time a dust grain remains heated around 70 K by a cosmic-ray particle, estimated to be ${\sim}3.16{\times}10^{-19}$.  The number ``70 K'' is the peak temperature a dust grain can achieve from cosmic-ray heating balanced by evaporative cooling.

\subsubsection{Photodesorption}

UV photons impinging on dust grains can eject surface molecules into the gas phase.  The photodesorption rates have been experimentally measured or theoretically computed for ice species such as CO, H$_2$O, N$_2$, etc. \citep{Oberg2009a,Oberg2009,Potapov2019,Gonzalez2019}, and empirical formulae have been used for calculating their rates.

We use a formula of the form for the rate coefficient
\begin{equation}
  k = F Y_i \sigma \eta \left(1-e^{-n_\text{L}(i)}\right),
  \label{eqPhotoDesor}
\end{equation} 
where $F$ is the UV flux (in counts cm$^{-2}$ s$^{-1}$), $Y_i$ is the photo yield for species $i$, $\sigma$ is the dust cross section, and $\eta$ is the dust-to-gas number ratio.  The number of monolayers of species $i$ is:
\begin{equation}
  n_\text{L}(i) = \frac{y_i}{\eta \NSS},
  \label{eqMonolayer}
\end{equation} 
where $y_i$ is the abundance relative to hydrogen nucleus, and $\NSS$ is the number of surface reaction sites per grain (${\sim}10^6$ for a dust grain radius of 0.1 $\mu$m).
The yield $Y$ may depends on the dust temperature.  For ice species without experimental data, $Y$ is assumed to be $10^{-4}$ by default.  However, \refeq{eqPhotoDesor} does not accurately reflect all the complexities found in experiments and should be treated with care when photodesorption is expected to be important.

\subsubsection{Chemical desorption}

The heat released by exothermic surface reactions may be able to launch the products off the dust grains.  This non-thermal desorption mechanism is invoked to explain the observed relatively high abundances of species such as methanol in dark clouds \citep{Garrod2007}.  While this mechanism is based on the Rice-Ramsperger-Kassel (RRK) theory, in effect it amounts to adding a branch to each surface two-body reactions, with the products being the gas phase counterparts of the surface products.  Hence a new empirical parameter is required to describing the efficacy of chemical desorption, namely the branching ratio between the gas phase to surface channels.  \codename{} by default assumes a value of 0.05 for this parameter.

\subsubsection{Two-body reaction rates on dust grain surfaces}

For a surface reaction of the form $i+j\rightarrow\cdots$, its rate coefficient is
\begin{equation}
  k = -(k_{\text{mig},i} + k_{\text{mig},j}) \frac{p_{ij}}{\NSS\eta},
\end{equation} 
where $k_{\text{mig},i}$ is the surface migration rate of species $i$, $p_{ij}$ is the probability for $i$ and $j$ to cross over the reaction barrier when they arrive at the same surface site, $\eta$ is the dust-to-grain number ratio as defined in \refeq{eqEta}, and $\NSS$ is the number of reactive sites per grain.  The surface migration can be due to either thermal hopping:
\begin{equation}
  k_{\text{mig},i} = \nu_i\exp\left(-E_{\text{diff},i}/T_\text{d}\right)
\end{equation} 
or quantum tunneling (at very low temperatures):
\begin{equation}
  k_{\text{mig},i} = \nu_i\exp\left(-\frac{2l}{\hbar}\sqrt{2 m_i E_{\text{diff},i}}\right),
\end{equation} 
where $\nu_i$ is the characteristic vibration frequency as defined in \refeq{eqVibFreq}, $E_{\text{diff},i}$ is the energy barrier against surface migration for species $i$, which is usually taken to be a fraction (0.3--0.5) of the desorption energy, and $l$ is the width of the barrier.  The reaction probability $p_{ij}$ can be calculated similar to $k_{\text{mig},i}$, except that the factor $\nu_i$ before the exponential function is dropped, and the diffusion barrier is replaced by the reaction barrier.

As a side note, the formation of H$_2$ molecules on dust grain surfaces is treated as a normal two-body surface reaction.  This is different from some of the previous works, in which the H$_2$ formation rate is taken to be half of the adsorption rate of H atoms (see, e.g., \citealt{Holdship2017}).

\subsection{Three-phase model of gas-grain chemistry}

As already described in previous sections, molecules can be adsorbed on the surface of dust grains.  After being adsorbed, these molecules are able to migrate over the dust grain surface and may react with each other when two molecules meet at the same surface site.
Since a dust grain has a limited surface area, when many molecules have adsorbed on its surface, newly adsorbed ones will likely sit on top of another.  In this way, multiple ice layers will form on the dust grains.  In such a description of dust grain structure, usually only molecules on the topmost layer are assumed to be mobile and reactive, while molecules underneath (in the mantle) are ``frozen'' and nonreactive (however, see \citealt{Chang2014,Chang2016}).  When molecules in the topmost layer evaporate, the underneath layer will be revealed and becomes the new reactive layer.  This is the so-called three-phase model of interstellar chemistry, in which the three phases are: gas phase, dust grain surface, and dust grain mantle.

\codename{} has implemented the three-phase description of gas-grain chemistry, following the approach of \citet{Hasegawa1993} and \citet{Garrod2011} (for a detailed description, see also \citealt{Du2012}).
For each surface species (except for very weakly adsorbed species such as \ce{H2}) a new mantle species is included in the network.
For species $i$ with surface population per grain $\ns{i}$, its evolution equation is
\begin{equation}
  \begin{split}
  \frac{d \ns{i}}{dt} =& \sum_{j,k} k_{j,k} \ns{j}\ns{k} - \sum_{j} k_{ij}\ns{i}\ns{j} \\
  & + k_\text{ads}(i) \nga{i} - k_\text{ads} \ns{i}/\NSS \\
  & + k_\text{des} \nm{i} / \max(n_\text{M},n_\text{S}) - k_\text{des}(i)\ns{i},
  \end{split}
  \label{eqThreePhaseSurface}
\end{equation}
and the mantle population of $i$ evolves according to
\begin{equation}
  \begin{split}
    \frac{d \nm{i}}{dt} = k_\text{ads} \ns{i}/\NSS - k_\text{des} \nm{i} / \max(n_\text{M}, n_\text{S}).
  \end{split}
  \label{eqThreePhaseMantle}
\end{equation}
In \refeq{eqThreePhaseSurface} the two summations are over two-body reactions producing and consuming species $i$.  $k_\text{ads/des}(i)$ is the adsorption/desorption rate coefficient of $i$, and $\nga{i}$ is the gas phase population of $i$.
$k_\text{ads/des}$ is the total adsorption/desorption rate of all the species, $n_\text{S}$ is the total population of all surface species, $n_\text{M}$ is the total population of all mantle species, and $\NSS$ is the number of sites per grain.
The quantity ``population-per-grain'' $\ns{i}$ is related with abundance with respect to hydrogen nuclei $y_i$ by (see also \refeq{eqMonolayer})
\begin{equation}
  y_i = \ns{i} \eta,
\end{equation}
where $\eta$ is the dust-to-hydrogen number ratio defined before.

Note that the three-phase formulation is statistical in nature.  There are more accurate treatments such as the microscopic kinetic Monte Carlo method \citep{Cuppen2007,Garrod2013}, which tracks the location of each particle on the dust grain surface.

With the inclusion of three-phase chemistry, the code will run slower, not just because of the increase in the number of variables due to addition of mantle species, but also because of the increased non-linearity of \refeq{eqThreePhaseSurface} relative to \refeq{eqCanonical}.

\subsection{The workflow of \codename{}}

While \codename{} can be used as a stand-alone program, it is mainly intended to be used as a Python module.  Here we describe the basic workflow of \codename{} when used in this mode.
A minimal working example is shown in \reftab{tabExample}.

\begin{enumerate}
  \item Create the \code{ChemModel} object.  This is the central Python object to work with in \codename{}.  During the instantiation the user may provide the file names of the reaction network, the initial abundances, and the enthalpies (needed for reaction heat calculation).  These files could also be loaded later with the corresponding helper functions.  The user could also add reactions one by one using Python code, and this is the way to load reaction network stored in different formats.
  \item Set the physical parameters.  Similarly, this could be done with an input file, or using a helper function to set the value of each parameter by its name.  They could also be provided in the previous step.  Optionally, the user could add time-dependency for certain parameters by providing a lookup table.  Internally the code linearly interpolate the table to evaluate time-dependent parameters at arbitrary time.
  \item Do some ``boilerplate'' preparation work.  These include, e.g., sorting out reactions of different types, getting the elemental composition of each species and calculating their molecular masses and quantum mobilities, and calculating the heat of each reaction.  Such boilerplate work usually only need to be done once for each model, unless during some unusual investigations (e.g., changing the reaction network while the code is running).
  \item Setup the ODE solver.  This usually amounts to setting the solver ID, which is itself optional (could be useful when running in multi-process mode).  Details of the solver setup and preparation are concealed in the C++ code, which include constructing the sparse structure of the ODE system, allocation of the needed working memories, and assignment of flags defining the working modes of the solver.
  \item \label{itemStart}Feed the current time and abundances, and the required output time, to the solver, then wait for the output.
  \item Retrieve and store the output time (which could be different from the value provided in the previous step) and abundances.
  \item Terminate if the output time is equal to or greater than the targeted final evolution time; otherwise go to \refstep{itemStart}.
\end{enumerate}

\lstset{basicstyle=\ttfamily,breaklines=true}
\lstloadlanguages{Python}
\begin{table}
\centering
\begin{tabular}{c}
\hline\hline
\begin{minipage}{0.9\linewidth}
\begin{lstlisting}
import chempl
model = chempl.ChemModel(
    fReactions=b'network_file',
    fInitialAbundances=b'abundance_file',
    phy_params={
     b'Av': 10.0,
     b'G0_UV': 1.0,
     b'Ncol_H2': 1e22,
     b'T_dust': 10.0,
     b'T_gas': 10.0,
     b'chemdesorption_factor': 0.05,
     b'chi_cosmicray': 1.0,
     b'dust2gas_mass': 1e-2,
     b'dust_albedo': 0.6,
     b'dust_material_density': 2.0,
     b'dust_radius': 0.1e-4,
     b'dust_site_density': 1e15,
     b'dv_km_s': 1.0,
     b'mean_mol_weight': 1.4,
     b'n_gas': 2e4})
model.prepare()
model.set_solver()
t = 0.0
dt = 0.1
tRatio = 1.1
tMax = 3.15e10 # in seconds
nMax = 1000
store = {'t': [], 'y': []}
y = model.abundances
for i in range(nMax):
    t, y = model.update(y, t=t, dt=dt)
    store['t'].append(t)
    store['y'].append(y)
    if t >= tMax:
        break
    dt *= tRatio
\end{lstlisting}
\end{minipage}\\
\hline
\end{tabular}
\caption{A minimal working example of a chemical model written in Python using \codename{}.  Tested with Python~3.  The Python bytes objects (in the form of \texttt{b'***'}) instead of string objects are needed for interacting with internal C++ code.\label{tabExample}}
\end{table}

\section{Example models based on \CODENAME{}}
\label{secExample}

In this section we present a few example models calculated with \codename{}.  The chemical networks we use for these examples are the UMIST network, or extended from it.
The latest version of the UMIST network is released in 2012, hence it is called \rtlv{} \citep{McElroy2013}.  The default reaction network format supported by \codename{} is similar to the KIDA network \citep{Wakelam2012}, instead of the colon-separated format of \rtlv{}, and can support any format with a simple conversion.

\subsection{Dark cloud models}

We first run a model with physical conditions relevant for dark clouds.  The physical parameters are listed in \reftab{tabPhyParamsDarkCloud}.
The initial abundances are listed in \reftab{tabIniDarkCloud}.
The network for this model is UMIST \rtlv{}, extended by only adding the \ce{H2} formation reaction on dust grain surface and the adsorption and evaporation of H and \ce{H2}, and no other adsorption/desorption processes and surface reactions are included.

\begin{figure}
  \centering
  \includegraphics[width=\linewidth]{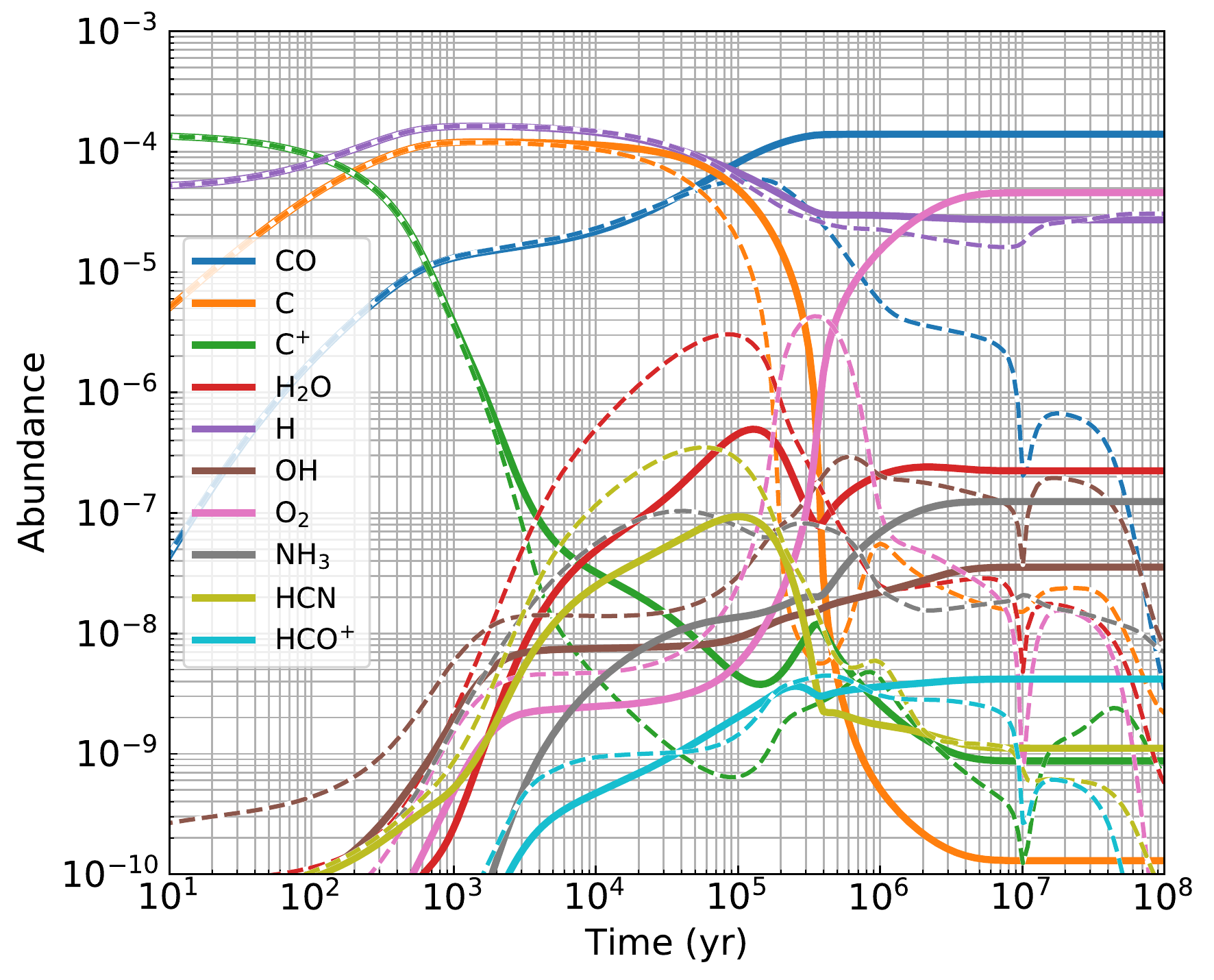}
  \caption{Example results from a basic dark cloud model.  Solid curves are calculated without any dust grain chemistry except \ce{H2} formation, and dashed curves with the same color are based on a full three-phase chemical network.}
  \label{figDarkCloudModel}
\end{figure}

The code takes 4--5 seconds of CPU time\footnote{All the example models reported in this paper are computed with a MacBook Pro (15-inch, 2017), with a processor configuration of ``3.1 GHz Quad-Core Intel Core i7''.} to reach $10^8$ yr.
The resulting evolution of a few common species are plotted as solid curves in \reffig{figDarkCloudModel}.
As can be seen, the system reaches steady state at ${\sim}10^7$ yr, after which carbon mainly resides in CO, while oxygen mainly resides in CO and \ce{O2}.
However, this model is not realistic at late times (${\gtrsim}10^6$ yr), because at the low temperature for this model, CO should freeze out onto dust grains.  So this model should be only viewed as a code benchmark and it does not accurately describe the actual dark cloud chemical processes.

So we run another model with the same initial abundances and physical parameters, but with a full three-phase chemical network.  Nonthermal desorption \citep{Garrod2007} by the heat released in surface reactions is also included.
With this added complexity, the code takes $\sim$15 seconds of CPU time to reach $10^8$ yr.
The results are plotted as dashed curves in \reffig{figDarkCloudModel}.
For species such as CO and C, their early-time (${\lesssim}10^4$ yr) evolution from the two models coincide.
This is because their abundances are mainly determined by gas-phase reactions at early times.
For species such as \ce{H2O}, OH, \ce{O2}, and \ce{NH3}, their evolution tracks from the two models depart away from each other from the very beginning.  This is because they are efficiently formed on the dust grain surface, and the inclusion of surface reactions (together with nonthermal desorption) enhances their abundances from the very beginning.

\begin{table}
\centering
\begin{tabular}{ll}
\hline\hline
$A_\text{v}$  &  10 \\
$G_0$  &  1 \\
$N_{\text{H}_2}$  &  10$^{22}$ cm$^{-2}$ \\
$n_\text{H}$  &  $2{\times}10^4$ cm$^{-3}$ \\
$T_\text{dust}$  &  10 K \\
$T_\text{gas}$  &  10 K \\
$\chi_\text{CR}$  &  $1.3{\times}10^{-17}$ s$^{-1}$ \\
Dust-to-gas mass ratio  &  0.01 \\
Dust albedo  &  0.6 \\
Dust material density  &  2.0 g cm$^{-3}$ \\
Dust grain radius  &  0.1 $\mu$m \\
Dust site density  &  $10^{15}$ cm$^{-2}$ \\
Chemical desorption efficiency  &  0.05 \\
\hline
\end{tabular}
\caption{Physical parameters used in the example dark cloud model.}
\label{tabPhyParamsDarkCloud}
\end{table}

\begin{table}
\centering
\begin{tabular}{ll}
\hline\hline
H$_{2}$  & $0.5$ \\
H  & $5{\times} 10^{-5}$ \\
He  & $9{\times} 10^{-2}$ \\
C$^{+}$  & $1.4{\times} 10^{-4}$ \\
N  & $7.5{\times} 10^{-5}$ \\
O  & $3.2{\times} 10^{-4}$ \\
S$^{+}$  & $8{\times} 10^{-8}$ \\
Si$^{+}$  & $8{\times} 10^{-9}$ \\
Fe$^{+}$  & $3{\times} 10^{-9}$ \\
F  & $2{\times} 10^{-8}$ \\
Cl$^{+}$  & $4{\times} 10^{-9}$ \\
Mg$^{+}$  & $7{\times} 10^{-9}$ \\
Na$^{+}$  & $2{\times} 10^{-9}$ \\
P$^{+}$  & $3{\times} 10^{-9}$ \\
E$^{-}$  & $1.40107{\times} 10^{-4}$ \\
\hline
\end{tabular}
\caption{Initial abundances used in the example dark cloud model.}
\label{tabIniDarkCloud}
\end{table}

\subsection{Time-dependent physical conditions}

Chemical models in which physical parameters (most frequently, density and temperature) change with time have been used in the study of warm-up phase of hot cores \citep{Garrod2006}.
Technically, this time-dependency could be implemented with a series of models with constant physical parameters, and the parameters only gradually change from one model to another.   In the series of models the final abundances of a previous model is used as initial condition of the next.  However, such an implementation would be computationally inefficient, because the ODE solver would have to reinitialize itself for each individual model, which is a slow process.
So it is necessary to add explicit time-dependency directly in a model, specifically, in the right hand side of \refeq{eqGeneric} or \refeq{eqCanonical}.

\subsubsection{Continuity of reaction rates}
\label{secContReacRates}

In the \rtlv{} network each reaction has associated a temperature range in which the rate parameters are most reliable.  Some reactions have two temperature ranges, which means different set of rate parameters must be chosen at different temperatures.  This could cause problems when temperature varies as a function of time: at the boundary of two temperature ranges, the reaction rates calculated from two versions of rate parameters are not necessarily continuous, which may confuse the ODE solver and the solution will become unreliable (or the solver will stop working and no solution can be obtained after the boundary of a temperature range is reached).

\codename{} ensures continuity (smoothness actually) by joining the rates calculated with two sets of parameters using a weighting function of the form
\begin{equation}
  g(x) = \frac{1}{1+e^{\pm(x-x_0)/w}},
\end{equation}
where $x_0$ is the interval edge, and $w>0$ is the separation from one interval to another.
The $\pm$ sign takes ``$+$'' when $x_0$ is the right edge, and takes ``$-$'' when $x_0$ is the left edge.
The function $g(x)$ has the property that for $x<x_0$, it quickly approaches unity, while for $x>x_0$, it quickly approaches zero.  The choice of weighting function is arbitrary, in the sense that a weighting scheme is allowed as far as the transition across the boundary is smooth and without any artificial bumps.

\begin{figure*}
  \centering
  \includegraphics[width=\linewidth]{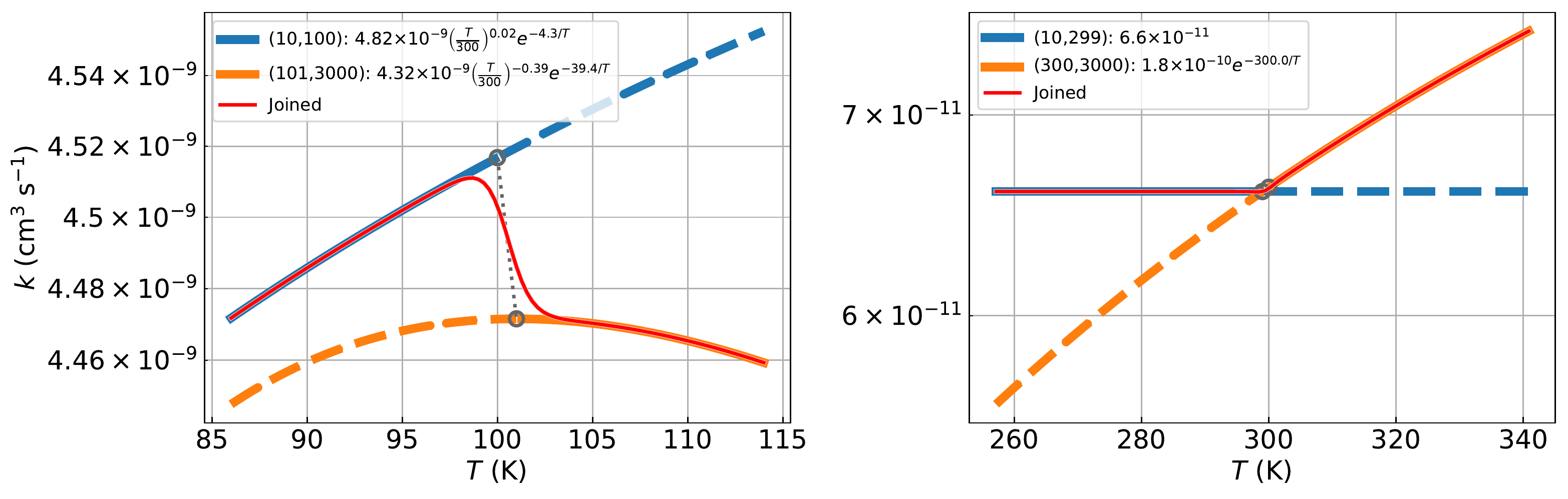}
  \caption{Examples showing how the rate coefficients for reactions with two temperature-range-of-applicability in the RATE12 database are calculated to ensure continuity and smoothness with respect to temperature.  The left panel corresponds to reaction \ce{H- + H -> H2 + E-}, and the right panel corresponds to \ce{NH + O -> NO + H}.  Blue and orange curves are the rate coefficients calculated with expressions for each temperature range as shown in the legends (dashed parts are outside the range of applicability).  The red curves are the smoothed rates joining the two used in the code.}
  \label{figSmoothRATE12}
\end{figure*}

\reffig{figSmoothRATE12} shows the rate coefficients of two reactions as a function of temperature.  As can be seen from the left panel, without a treatment such as the one described here, the rate coefficient would jump from one value to another when the temperature only slightly varies at the boundary (from 100 to 101 K).  This may halt the ODE solver since it will have to use exceedingly small time steps (since temperature is a function of time in time-dependent models).
With a continuous function, the ODE solver will be able to integrate much easier.

\begin{figure}
  \centering
  \includegraphics[width=\linewidth]{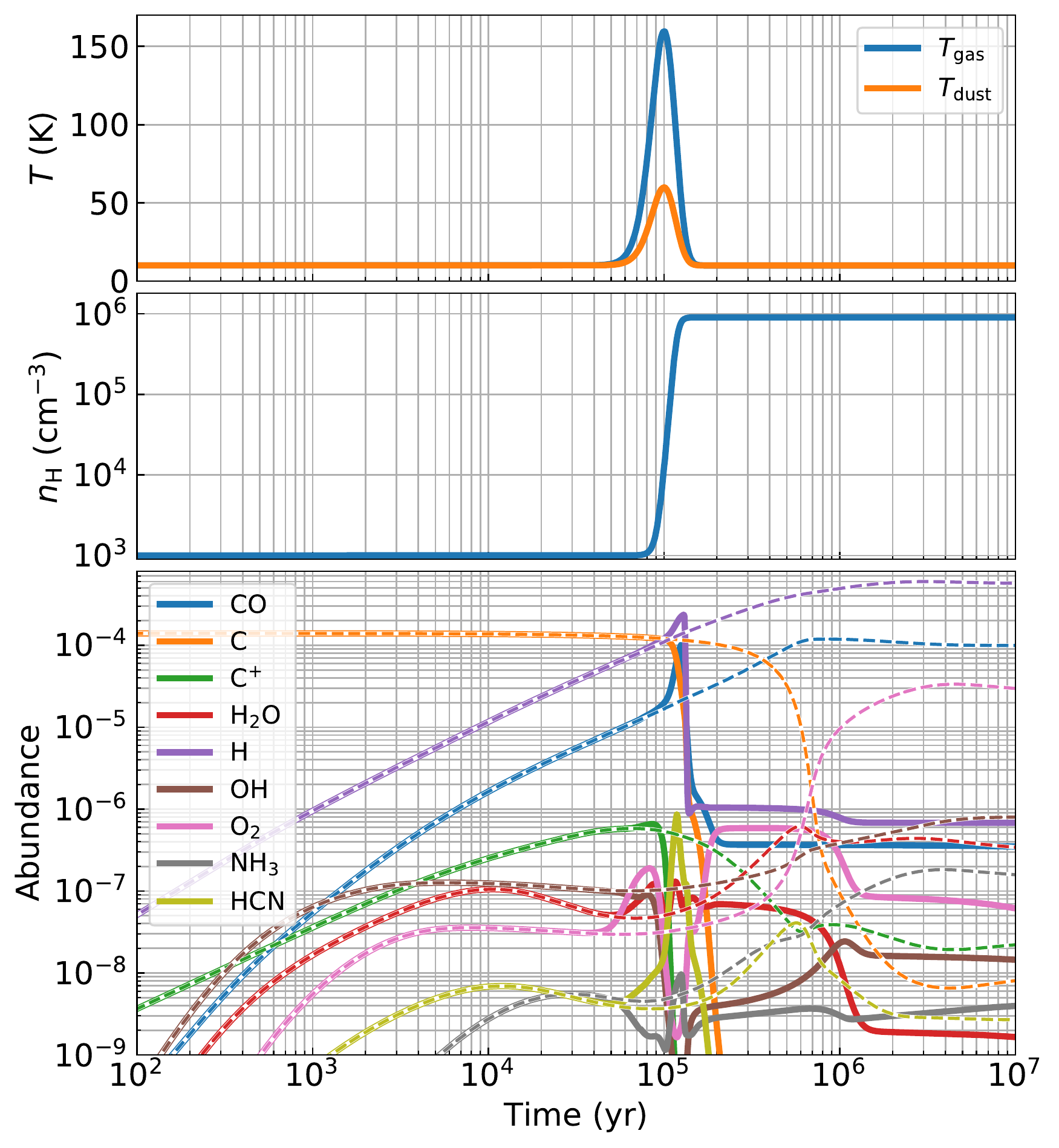}
  \caption{Results from a toy model with time-dependent temperature and density.  The top and middle panels show the gas and dust temperature and gas density as a function of time.  Solid curves in the bottom panel show the corresponding evolution of abundances of different species, while dashed curves in that panel are calculated with temperature and density kept constant at their initial values.}
  \label{figTimeDependent}
\end{figure}

\begin{figure}
  \centering
  \includegraphics[width=\linewidth]{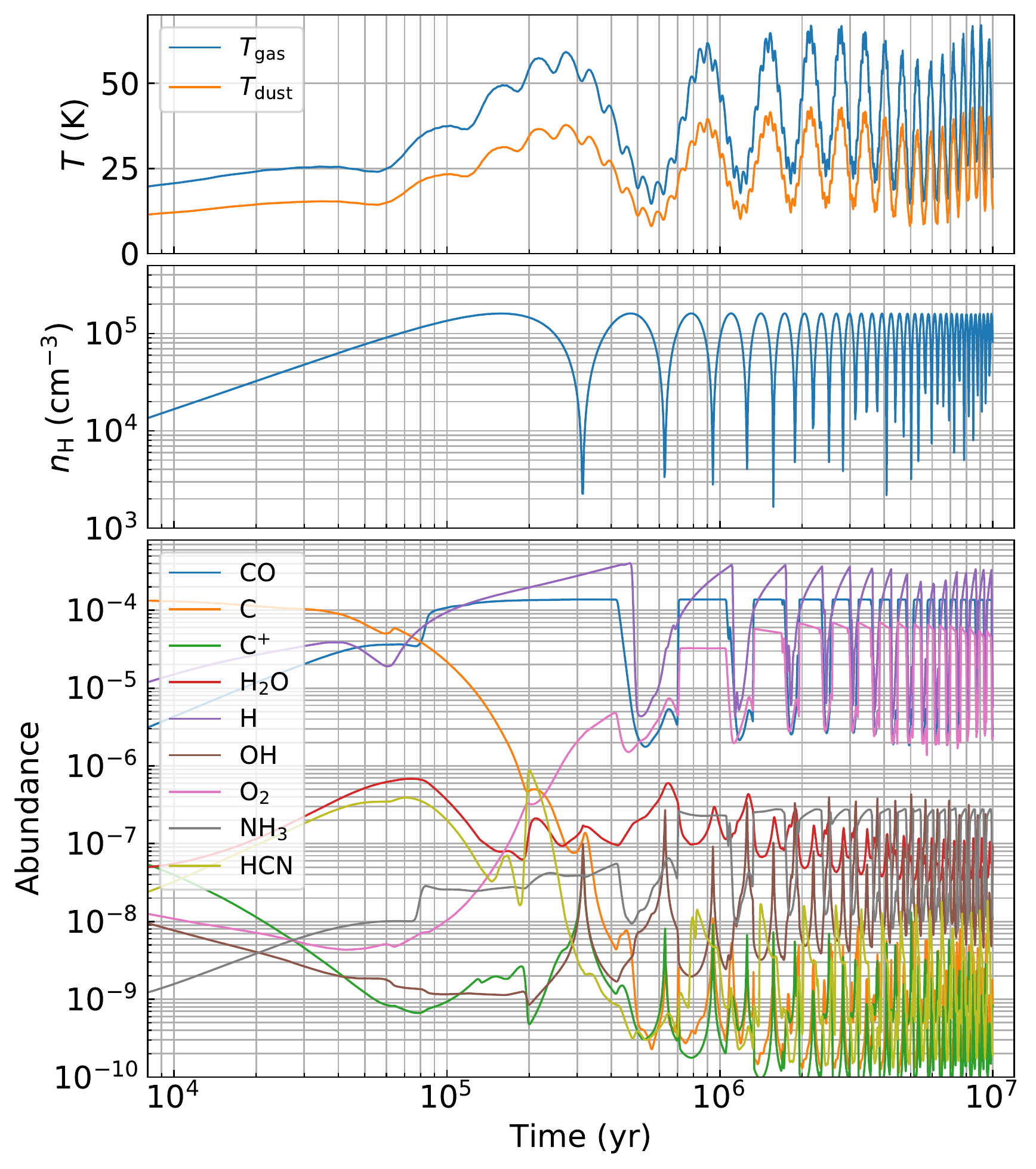}
  \caption{Similar to \reffig{figTimeDependent}.  Results from a toy model with ``wavy'' and spiky time-dependent temperature and density.}
  \label{figTimeDependentRand}
\end{figure}

\reffig{figTimeDependent} and \reffig{figTimeDependentRand} show two example models with time-dependent temperature and density.
The reaction network used for them are \rtlv{} with the addition of adsorption and desorption reactions, while no surface reactions beyond the formation of \ce{H2} are included.
The time evolution of temperature and density in the two models are arbitrarily designed, while keeping in mind that they may mimic realistic processes occurring in certain interstellar environments.
The first model simulates a compressing and heating event, which takes about 40 CPU seconds to reach $10^7$ yr, while the second one simulates a ``wavy'' or oscillating temperature and density profile, and takes about 16 CPU minutes to reach $10^7$ yr.
The abrupt oscillation of temperature and density, especially at late times, limits the step size used internally in the ODE solver and significantly slows down the integration speed.

Interestingly, the evolution curves plotted in the bottom panel of \reffig{figTimeDependentRand} demonstrate two types of behaviors: those which follow temperature (CO, \ce{O2}, and \ce{NH3}), and those which follow density (\ce{H2O}, \ce{OH}).
For species of the first type, their abundances are mostly controlled by desorption, hence their abundances jumps when the dust temperature reaches their desorption temperature, and stays flat after that.  For species of the second type, either they are not able to evaporate at dust temperatures $\lesssim$40 K (the highest dust temperature in the model), and/or their abundances are mostly determined by gas phase reactions, and their abundances tend to follow the evolution of density.  The behavior of C and \ce{C+} are more complex.  They seem to follow both density and the rise and drop of temperature, which is because their abundances are affected by gas phase reactions which involves species that can evaporate at dust temperature less than 40 K.

\subsection{Circumstellar envelope}
\label{secCircumstellar}

Since we largely use the UMIST \rtlv{} network as a starting point for \codename{}, we also run a model similar to the ``carbon-rich circumstellar envelope model'' presented in the \rtlv{} paper \citep{McElroy2013}.
This may be viewed as a non-strict benchmark, because a strict one is possible only if the exact parameters (including physical constants) used in their model are adopted.  Here we try to use the same input as them, though an exact match is not intended for.
The initial abundances and density profile are taken from Table 6 of \citet{McElroy2013}, and the temperature profile is taken from \citet{Cordiner2009}.
The self-shielding of CO from photodissociation is treated using the approach of \citet{Morris1983}, which is also the same as in \citet{McElroy2013}.
While this model calculates the spatial structure of a circumstellar envelope around a carbon-rich AGB star, however, since the gas in the envelope are launched from very close to the central star with a fully-specified velocity profile, there is an one-to-one mapping between time and radius.  Thus the model can be viewed as one in which physical parameters vary with time, which is how we implement it.

\begin{figure}
  \centering
  \includegraphics[width=\linewidth]{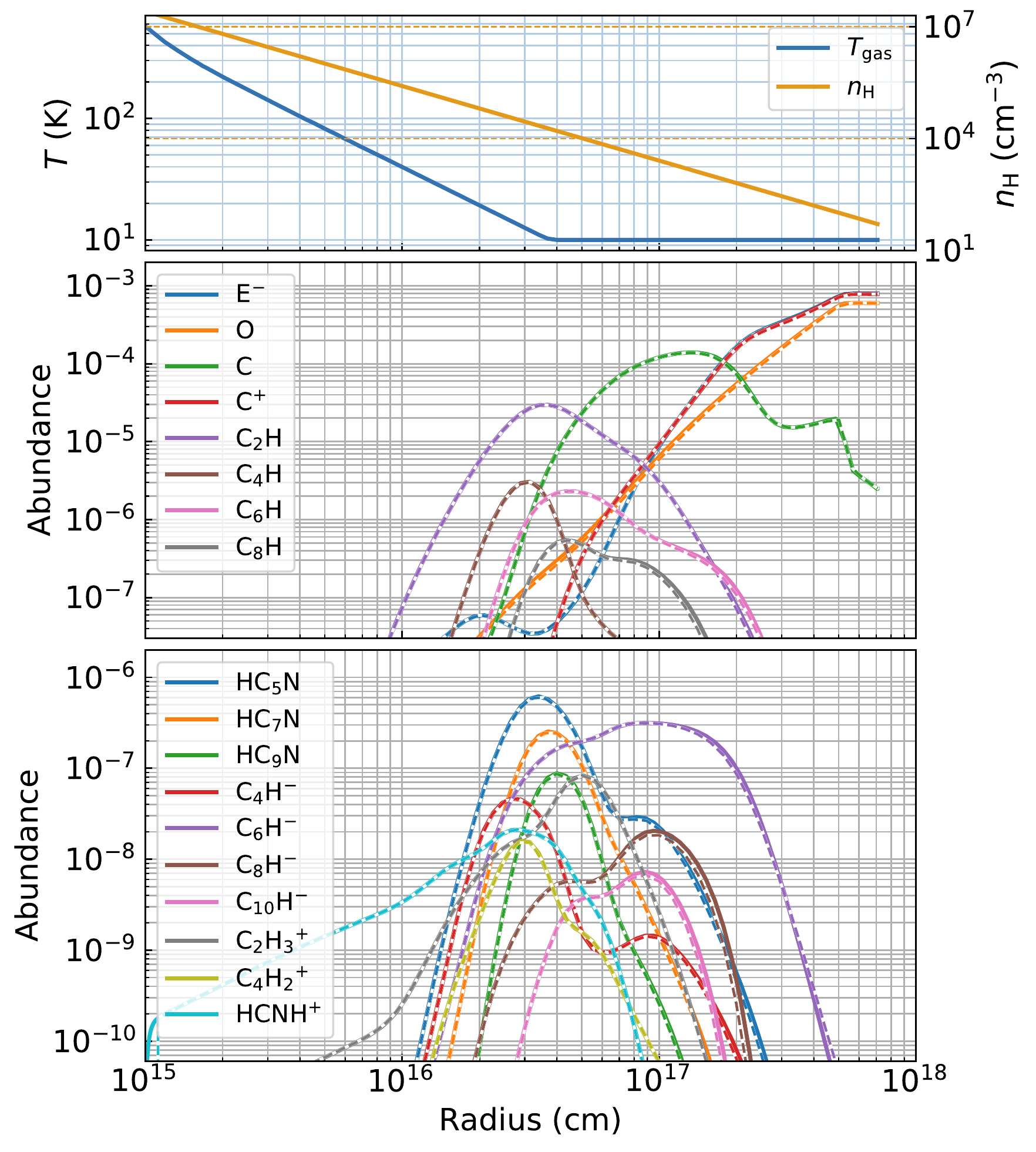}
  \caption{A simple circumstellar envelope model.  The solid curves in the lower two panels are calculated with \codename{}, while the dashed curves are taken from Fig.~2 of \citet{McElroy2013} using data file generated with their code.  The top panel shows the adopted temperature (blue) and density (orange) profile.}
  \label{figEnvelope}
\end{figure}

The results of the calculation with \codename{} are shown as solid curves in \reffig{figEnvelope}, overlaid with data from \citet{McElroy2013} shown as dashed curves.  It is clear that the two calculations match very well: two versions of most of the curves are almost identical within numerical errors.  This close match means our implementation of the envelope structure with \codename{} is correct (although there might be some fine details that are not reproduced in our model).

\subsection{Spatially dependent physical conditions: a toy PDR chemical model}

As a final example, we create a model that simulates a plane-parallel photodissociation region.  The density is taken to be uniformly $10^3$~cm$^{-3}$, the $G_0$ parameter of UV field at the edge is set to 100, while the dust temperature profile is calculated using equation (9.18) of \citet{Tielens2005}, and the gas temperature is assumed to be uniformly 100~K.

The calculation has to proceed from the surface layer ($A_\text{V}\sim 0$) to the deeper layers to take into account the self-shielding of \ce{H2} and \ce{CO}.  Here the CO self-shielding is approximated using the same approach as in \refsec{secCircumstellar}, while the \ce{H2} self-shielding is implemented following \citet{Draine1996}.  More accurate treatment of self and mutual shielding can also be implemented \citep{Visser2009,Li2013,Heays2017}.

\begin{figure}
  \centering
  \includegraphics[width=\linewidth]{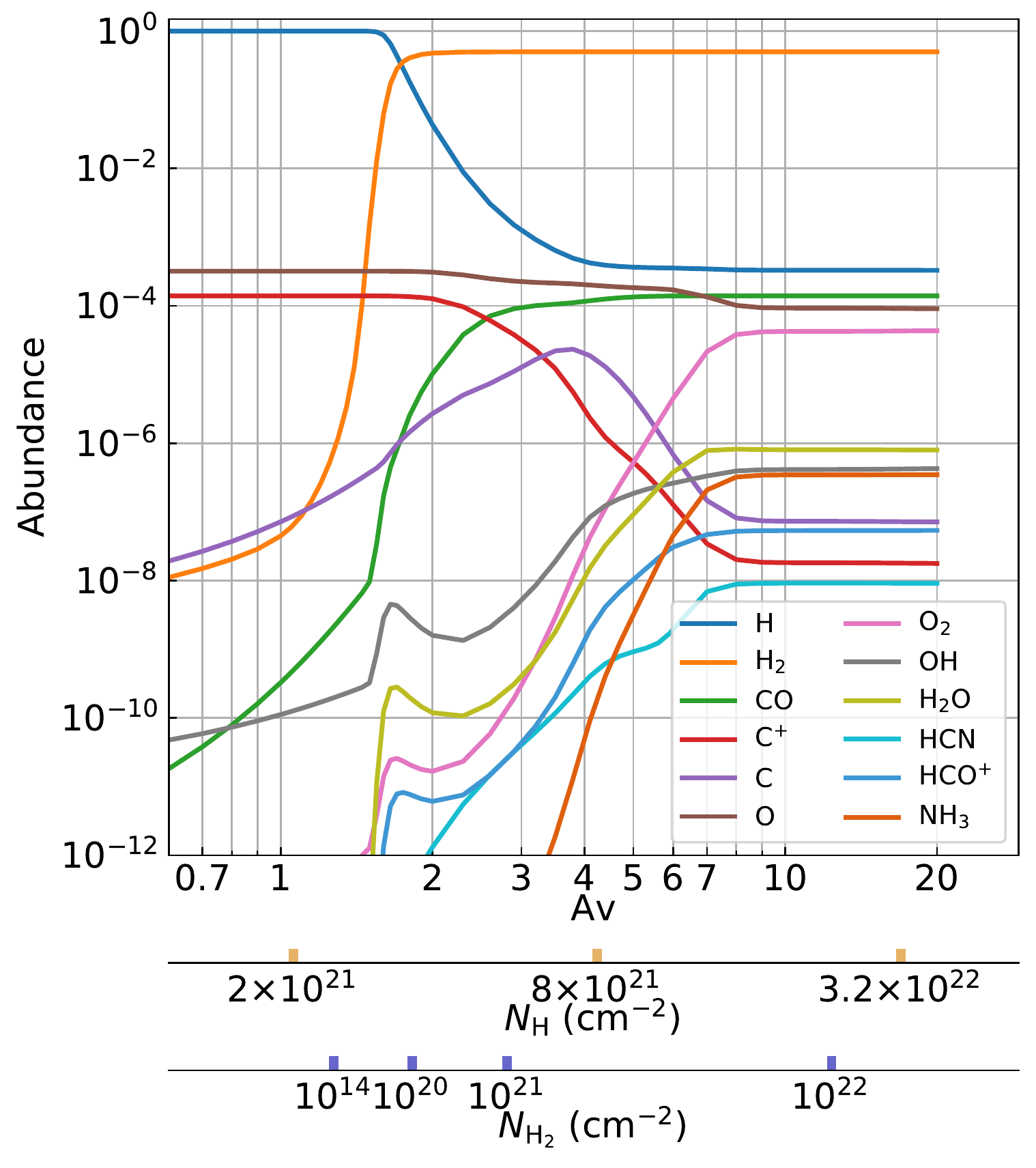}
  \caption{A simple PDR model, which shows the abundances of different species as a function of depth (visual extinction $A_\text{V}$ or column density of hydrogen nucleus $N_\text{H}$).  The column density of molecular hydrogen at each depth is marked in the bottom horizontal axis.}
  \label{figPDR}
\end{figure}

The resulting abundances of different molecules are shown as a function of depth ($A_\text{V}$) and column density are shown in \reffig{figPDR}.  As is commonly known, the H/\ce{H2} transition occurs at $A_\text{V}\sim 1-2$, while the \ce{C+}/C/CO transition occurs at $A_\text{V}\sim 2-4$ (the exact transition locations depend on the density profile and the intensity of the external UV field).

\subsection{Model-level parallelism}

As shown before, when the physical parameters are kept constant, \codename{} runs relatively fast.
So it is not necessary to parallelize each individual model.  However, when doing a parameter study, in which many models with different configurations need to be computed, parallelization becomes helpful.  In the case of \codename{}, the current parallelization approach is to make use of the Python built-in ``multiprocessing'' module.  With this module, it is possible to run multiple independent models at the same time, which can accelerate the computational speed by a factor depending on the number of CPU cores of the system.  We call this approach of parallelization ``model-level'' parallelism\footnote{Since there is no interdependence between the models, this type of parallelism is considered to be ``embarrassingly parallel''.}.

When the physical parameters explicitly change with time, especially when the time variation is oscillatory and spiky, as shown in the example model in \reffig{figTimeDependentRand}, the computational speed will be significantly reduced.  We don't expect an easy solution to alleviate the computation burden here, because the problem is inherently difficult.
In principle, one could parallelize the matrix inversion (which is frequently carried out in the solver) in the ODE solver to speed it up, however the effectiveness of doing this remains to be tested.

\section{Conclusions}
\label{secConclusion}

In this paper we introduce a new code named \codename{} for astrochemical study.  In comparison with other codes in the field, \codename{} emphasizes interactivity and ``playability''.  While \codename{} can be used as a traditional standalone command line program or called as a subroutine within a bigger simulation, it is mainly meant to be used as a Python module.  In the recent years the Python programming language and its ecosystem have become prevalent in astronomical community, and \codename{} is joining this trend.

\codename{} is not a completed project.  It is still evolving.  We expect in the future it will keep improving both in performance and in ease-of-use, as well as in incorporating more chemical and physical mechanisms (for example solving the temperature in tandem with chemical abundances, as done in \citealt{Du2014}).
One potential feature that might be handy is for the user to be able to choose a different ODE solver.
It may also be useful for the chemical solver to be able to switch between the rate equation approach and the Monte Carlo \citep{Vasyunin2009,Chang2016}, the moment equation \citep{Du2011}, or the modified-rate-equation approaches \citep{Caselli1998,Garrod2008}.
This will facilitate the study of the fundamental issue of the stochasticity of grain surface reactions.

\section*{Acknowledgements}

We thank the anonymous referee for making this paper a better read.
This work is partially funded by the National Natural Science Foundation of China (Project number 11873094 and 11873097).  F.~Du is financially supported by the Hundred Talents Program of Chinese Academy of Sciences.
This work makes heavy usage of Cython, Jupyter Notebook, Numpy, Matplotlib, etc{.} as delivered in the Anaconda platform.  The ODEPACK solver collection and the GCC compiler collection are also instrumental for this work.

\bibliographystyle{raa}
\bibliography{ms2020-0272}

\label{lastpage}

\end{document}